\documentclass[seceq]{ptptex}

\usepackage{graphicx}
\usepackage{wrapft}

\newcommand{\ps}{p\hspace{-5pt}/}
\newcommand{\as}{A\hspace{-6pt}/}

\title{Chromomagnetic Instability and Gluonic Phase\\
in Dense Neutral Quark Matter}
\author{Osamu \textsc{Kiriyama}}

\inst{Institut f\"ur Theoretische Physik, J.W. Goethe-Universit\"at,\\
D-60438 Frankfurt am Main, Germany}

\abst{
We discuss the chromomagnetic instability problem in the 2SC/g2SC phases 
and explore the phase structure of the gluonic phase 
which is a candidate for the ground state of dense neutral quark matter.
}
\begin{document}
\maketitle

\section{Introduction}
The phase structure of hot and/or dense QCD 
is one of the most exciting current topics 
in the field of strong interactions. 
In particular, the properties of cold and dense quark matter 
are of great interest in astrophysics and cosmology; 
it is now widely accepted that, at moderate densities of relevance for 
the interior of compact stars, quark matter is a color superconductor 
and has a rich phase structure 
with important implications for compact star physics 
(for a recent review, see Ref. \citen{CSC}).

Bulk matter in the interior of compact stars 
should be color and electrically neutral and be in $\beta$-equilibrium. 
In the two-flavor case, these conditions separate 
the Fermi momenta of up and down quarks and, 
as a consequence, the ordinary BCS state (2SC) is not always 
energetically favored over other unconventional states. 
The possibilities include crystalline 
color superconductivity \cite{LOFF,ABR} and 
gapless color superconductivity (g2SC) \cite{Shovkovy2003}. 
However, the 2SC/g2SC phases suffer from a chromomagnetic instability, 
indicated by imaginary Meissner masses of some gluons \cite{Huang2004}. 
The instability related to gluons of color 4--7 
occurs when the ratio of the 2SC gap 
over the chemical potential mismatch, $\Delta/\delta\mu$, 
decreases below a value $\sqrt{2}$. 
Resolving the chromomagnetic instability 
and clarifying the nature of true ground state of dense quark matter 
are central issues in the study of color superconductivity 
\cite{Gorbar2005,Gorbar2005b,Fukush2006,
KRS2006,Kiri2006,HJZ,GR,Hashimoto2007}. 

We will describe the results of recent studies 
of a chromomagnetic instability and 
a gluonic phase \cite{Gorbar2005} (gluonic cylindrical phase II) 
in dense neutral quark matter.

\section{Model, Formalism and Numerical Results}
In order to study the chromomagnetic instability 
and phases with gluonic vector condensates, 
we use a gauged Nambu.Jona-Lasinio (NJL) model with massless
up and down quarks:
\begin{eqnarray}
{\cal L}=\bar{\psi}(iD\hspace{-7pt}/+\hat{\mu}\gamma^0)\psi
+G_D(\bar{\psi}i\gamma_5\varepsilon\epsilon^bC\bar{\psi}^T)
(\psi Ci\gamma_5\varepsilon\epsilon^b\psi)
-\frac{1}{4}F_{\mu\nu}^{a}F^{a\mu\nu},
\end{eqnarray}
where the quark field $\psi$ carries flavor ($i,j=1,\ldots N_f$
with $N_f=2$) and color ($\alpha,\beta=1,\ldots N_c$ with $N_c=3$)
indices, $C$ is the charge conjugation matrix;
$(\varepsilon)^{ik}=\varepsilon^{ik}$ and
$(\epsilon^b)^{\alpha\beta}=\epsilon^{b\alpha\beta}$
are the antisymmetric tensors in flavor and color spaces,
respectively. The diquark coupling strength
in the scalar color-antitriplet channel
is denoted by $G_D$. The covariant derivative and the field
strength tensor are defined as
\begin{eqnarray}
D_{\mu} = \partial_{\mu}-igA_{\mu}^{a}T^{a},
~F_{\mu\nu}^{a} = \partial_{\mu}A_{\nu}^{a}-\partial_{\nu}A_{\mu}^{a}
+gf^{abc}A_{\mu}^{b}A_{\nu}^{c}.
\end{eqnarray}

In NJL-type models without dynamic gauge fields,
one has to introduce color
and electric chemical potentials by hand 
to ensure color- and electric-charge neutrality.
In $\beta$-equilibrated neutral two-flavor quark matter,
the elements of the diagonal matrix of
quark chemical potentials $\hat{\mu}$ are given by
\begin{eqnarray}
&&\mu_{ur}=\mu_{ug}=\bar{\mu}-\delta\mu,
~\mu_{dr}=\mu_{dg}=\bar{\mu}+\delta\mu,\nonumber\\
&&\mu_{ub}=\bar{\mu}-\delta\mu-\mu_8,
~\mu_{db}=\bar{\mu}+\delta\mu-\mu_8,
\end{eqnarray}
with $\bar{\mu}=\mu-\delta\mu/3+\mu_8/3$ and $\delta\mu=\mu_e/2$.

In the mean-field approximation, the effective potential 
with constant gluonic vector condensates is given by
\begin{eqnarray}
\Omega_R=\Omega(\Delta,\delta\mu,\mu_a,\vec{A}^a;\mu,T)
-\Omega(0,0,0,\vec{A}^a;0,0),
\end{eqnarray}
where
\begin{eqnarray}
\Omega &=& \frac{\Delta^2}{4G_D}
-\frac{1}{2}\int\frac{d^4p}{(2\pi)^4i}{\rm Tr}\ln S^{-1}(p) 
-\frac{\mu_e^4}{12\pi^2}
+\frac{g^2}{4}f^{abc}f^{ade}A_{\mu}^bA_{\nu}^cA^{d\mu}A^{e\nu}.
\end{eqnarray}
Here $S^{-1}(p)$ is the inverse full quark propagator 
in Nambu-Gor'kov space,
\begin{eqnarray}
S^{-1}(p)=\left(
\begin{array}{cc}
\ps+(\bar{\mu}-\delta\mu\tau^3)\gamma^0+g\as^aT^a
& -i\varepsilon\epsilon^b\gamma_5\Delta\\
-i\varepsilon\epsilon^b\gamma_5\Delta
& \ps-(\bar{\mu}-\delta\mu\tau^3)\gamma^0-g\as^{a}T^{aT}\end{array}
\right),\label{eqn:propagator}
\end{eqnarray}
where $\tau^3=\mbox{diag}(1,-1)$ is a matrix in flavor space. 
Following the usual convention, 
we have chosen the diquark condensate to point 
in the third direction in color space. 
In order to evaluate loop diagrams we use a three-momentum cutoff
$\Lambda=653.3$ MeV throughout this paper.

\begin{figure}
\centerline{\includegraphics[width=0.75\textwidth,clip]{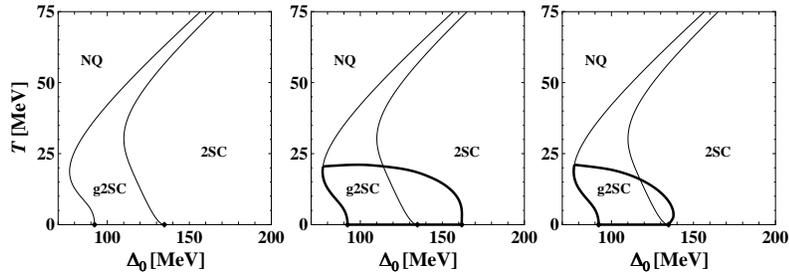}}
\caption{Left: The phase diagram of a neutral two-flavor
color superconductor 
in the plane of temperature and $\Delta_0$.\cite{Kiri2006}. 
At $T=0$, the g2SC phase exists in the window 
$92~{\rm MeV} < \Delta_0 < 134~{\rm MeV}$ 
and the 2SC window is given by $\Delta_0 > 134~{\rm MeV}$. 
The results are plotted for $\mu=400$ MeV. 
Middle: The same as the left panel, but the unstable region for
gluons 4--7 is depicted by the region enclosed by the thick solid line. 
At $T=0$, the g2SC phase and a part of the 2SC phase
($92~{\rm MeV} < \Delta_0 < 162~{\rm MeV}$) suffer from
the chromomagnetic instability. 
Right: The same as the left panel, but the unstable region for 
the 8th gluon is depicted by the region enclosed by the thick solid line.}
\label{Figure1}
\end{figure}

The chromomagnetic instability related to gluons 4--7 and 8 
can be viewed as tendencies toward 
the vector condensation of $\phi_{\mu}$ and 
$A_{\mu}^8$ \cite{Gorbar2005,Fukush2006,KRS2006,Kiri2006}, respectively, 
where $\phi_{\mu}=\sqrt{1/2}(A_{\mu}^4-iA_{\mu}^5,A_{\mu}^6-iA_{\mu}^7)^T$ 
is the doublet field with respect to 
the residual ${\rm SU}(2)_c$ symmetry in the 2SC/g2SC phases. 
Because of the ${\rm SU}(2)_c$ symmetry, 
we can choose $A_{\mu}^6$ without loss of generality. 

In Fig. \ref{Figure1}, we plotted the phase diagram 
of homogeneous neutral two-flavor quark matter 
in the plane of the 2SC gap at $\delta\mu=0$ ($\Delta_0$) 
and temperature ($T$). 
(The parameter $\Delta_0$ is essentially the diquark coupling strength. 
We neglect color chemical potentials throughout, 
because the present work remains qualitatively unaffected by them.) 
We calculated the squared Meissner masses of gluons 
from the second derivative of the effective potential 
with respect to $\langle \vec{A}^6 \rangle$ 
and $\langle \vec{A}^8 \rangle$ 
and mapped out the unstable regions for gluons 6 and 8 on the phase diagram. 
In the unstable regions, it is natural 
to suggest that the true ground state 
is given by the global minimum of the effective potential, 
including the gluonic vector condensation. 
Note that a $\langle \vec{A}^6 \rangle$-condensed phase 
is called gluonic phase and 
a $\langle \vec{A}^8 \rangle$-condensed phase 
is gauge equivalent to 
the (single plane-wave) LOFF phase.\cite{Gorbar2005,LOFF,ABR,Gorbar2005b}

\begin{wrapfigure}{r}{6.6cm}
\centerline{\includegraphics[width=6.3cm]{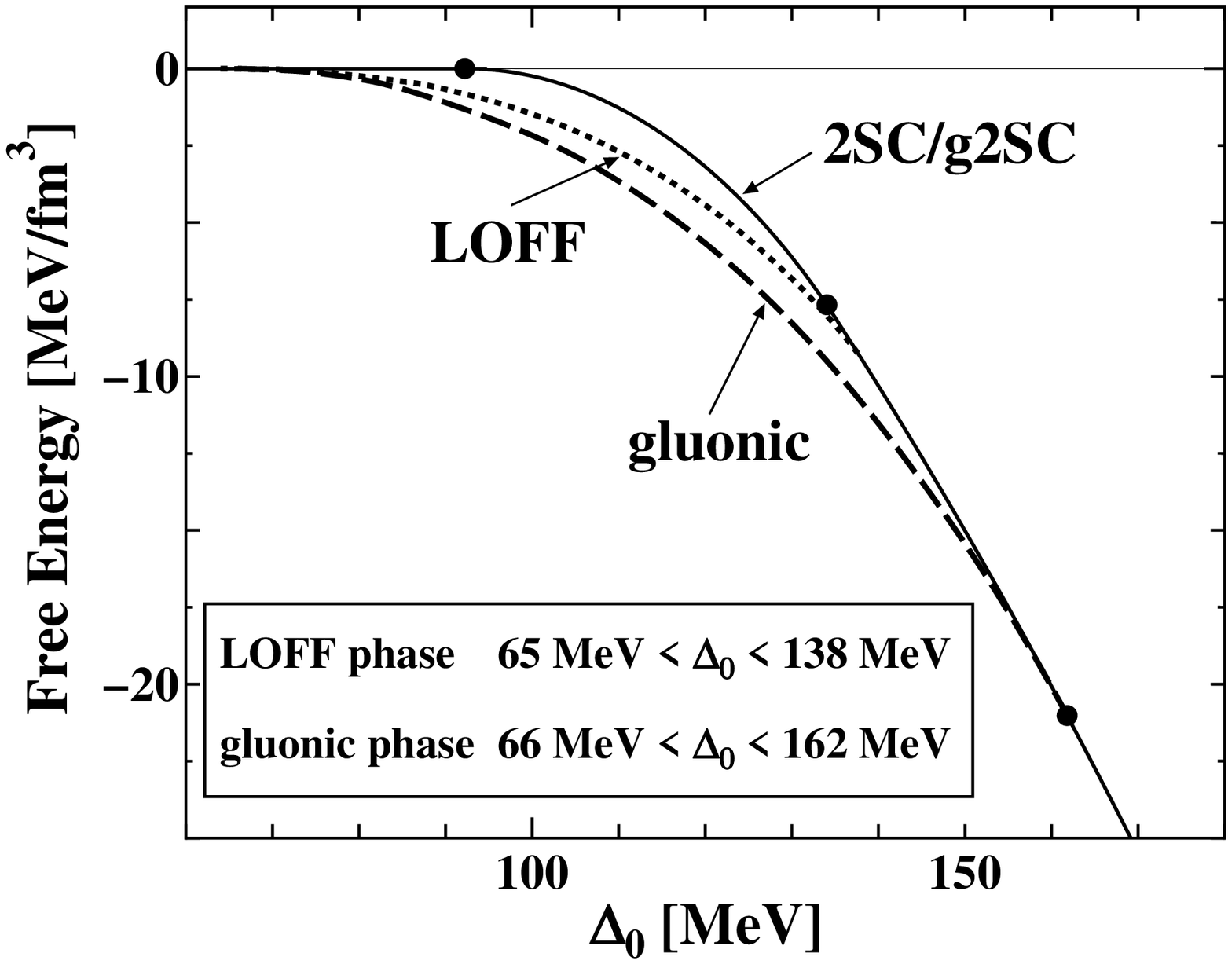}}
\caption{The free energies of the 2SC/g2SC
phase, the LOFF state, and the gluonic phase 
as a function of $\Delta_0$ (measured with respect to the unpaired NQ phase).
The three dots on the bold solid line denote the edge of the g2SC window with
the normal phase, the phase transition point between 2SC
and g2SC phases ($\Delta=\delta\mu$), and the critical point
of the chromomagnetic instability ($\Delta\simeq \sqrt{2}\delta\mu$).
The results are plotted for $\mu = 400~{\rm MeV}$ and $T=0$.}
\label{Figure2}
\end{wrapfigure}

\begin{figure}
\centerline{\includegraphics[width=0.8\textwidth,clip]{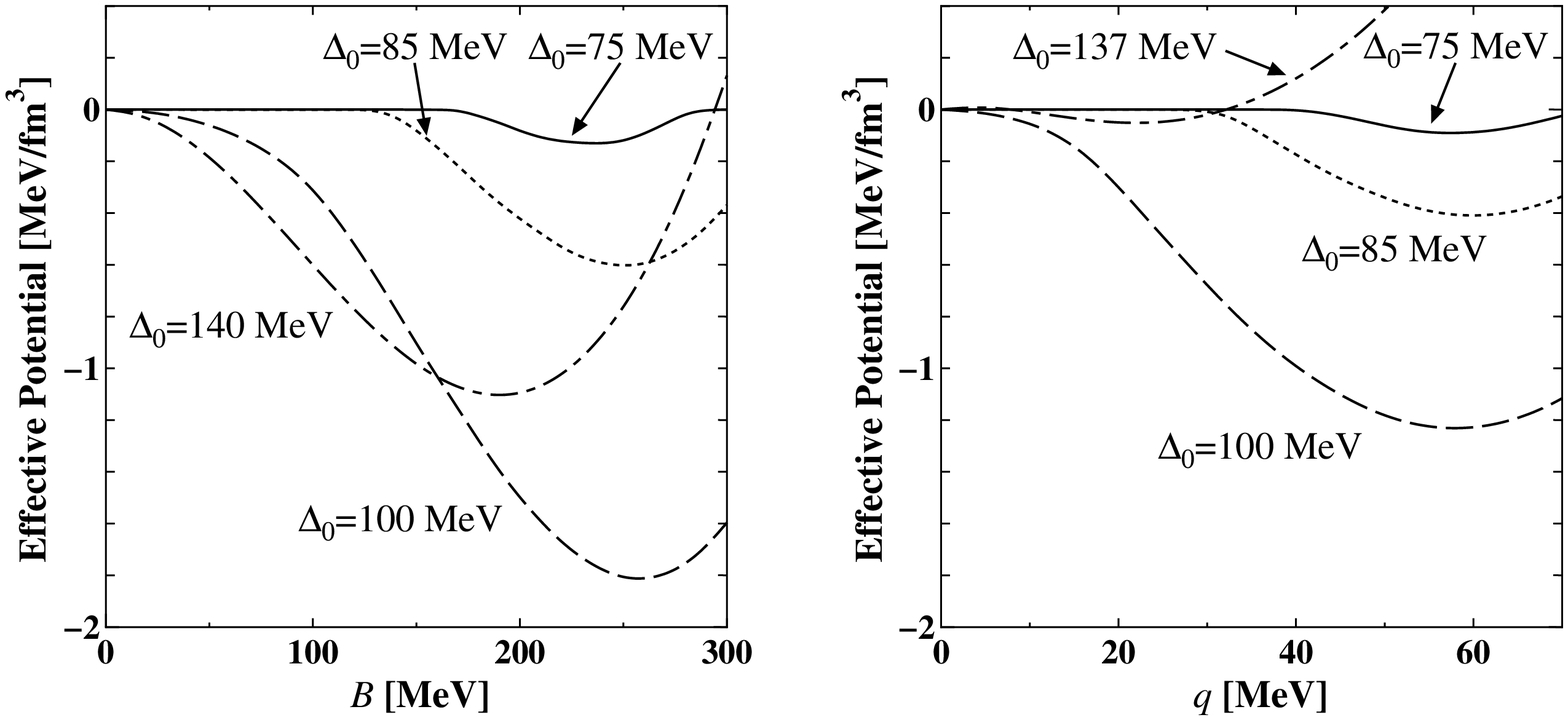}}
\caption{The effective potential of the gluonic (LOFF) phase as a function of 
$B=\langle gA_z^6 \rangle$ ($q=\langle gA_z^8 \rangle /(2\sqrt{3})$). 
Note that the effective potentials are calculated 
along the self-consistent solution of the gap equation for $\Delta$ 
and the neutrality consition for $\delta\mu$ 
and are measured with respect to the 2SC/g2SC/NQ phases at $B=0$ ($q=0$). 
The results are plotted for $\mu=400~{\rm MeV}$ and $T=0$.}
\label{Figure3}
\end{figure}

In order to find the gluonic/LOFF phases, 
we solved the gap equations for $\Delta$ 
and $\langle \vec{A}^a\rangle$ [$a\in(6,8)$] 
and the neutrality condition for $\delta\mu$ 
self-consistently 
and computed their free energies.\cite{Kiri2007} 
(This has been done first by Hashimoto 
and Miransky \cite{Hashimoto2007}.) 
We illustrate the comparison of the free energies 
of the gluonic/LOFF phases in Fig. \ref{Figure2}. 
The result is plotted for $\mu=400~{\rm MeV}$ and $T=0$ 
as a function of $\Delta_0$. The gluonic phase exists 
in the window $66~{\rm MeV} < \Delta_0 < 162~{\rm MeV}$
\footnote{The upper boundary of the gluonic (LOFF) phase is 
slightly larger than that of the unstable region 
for gluons 4--7 (8), because the transition from the gluonic (LOFF) phase 
to the stable 2SC phase is of first order.} 
and is energetically more favored than the LOFF phase 
in a wide range of the coupling strength. 
It is also quite interesting to note that the gluonic (LOFF) phase 
could be energetically more favored 
than the NQ phase for $66~{\rm MeV} < \Delta_0 < 92~{\rm MeV}$
($65~{\rm MeV} < \Delta_0 < 92~{\rm MeV}$), 
though the chromomagnetic instability does not exist 
in the NQ phase. By computing the effective potential 
as a function of the vector condensates 
along the self-consistent solution of the gap equation for $\Delta$ 
and the neutrality condition for $\delta\mu$, 
we revealed that it indeed happens 
in the weak-coupling regime (see Fig. \ref{Figure3}). 
In the following, we take account of the gluonic phase only 
and look at the phase diagram of two- and three-flavor quark matter.

\begin{wrapfigure}{r}{6.6cm}
\centerline{\includegraphics[width=6.3cm]{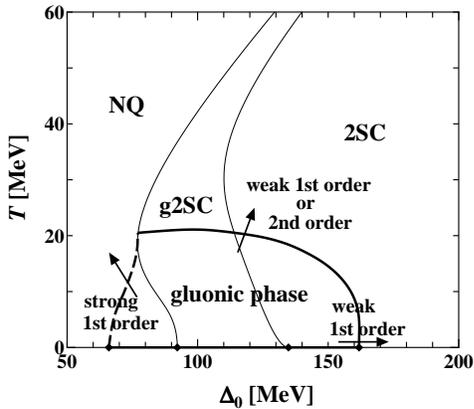}}
\caption{Schematic phase diagram of neutral two-flavor quark matter 
at $\mu=400$ MeV in $\Delta_0$-$T$ plane. 
In the region enclosed by the thick lines, 
the gluonic phase is energetically more favored
than the 2SC/g2SC/NQ phases.}
\label{Figure4}
\end{wrapfigure}

We extended our analysis to nonzero temperatures 
and obtained the schematic phase diagram 
shown in Fig. \ref{Figure4}. 
As expected, the low-temperature regions of the 2SC/g2SC phases 
are replaced by the gluonic phase. We also find that the transition 
between the gluonic phase and the 2SC/g2SC phases 
is of second order or weakly first order. 
Furthermore, the gluonic phase wins against 
a part of the NQ phase and undergoes a strong first-order 
phase transition into the NQ phase. 
(Here, it should be mentioned that a cutoff artifact appears 
in the gluon sector of the effective potential. 
Therefore it may be impossible to distinguish the second-order transition 
from the weak first-order transition. 
However, the cutoff artifact is not significantly large, 
hence we could calculate the phase diagram.)

\begin{figure}
\centerline{\includegraphics[width=0.75\textwidth,clip]{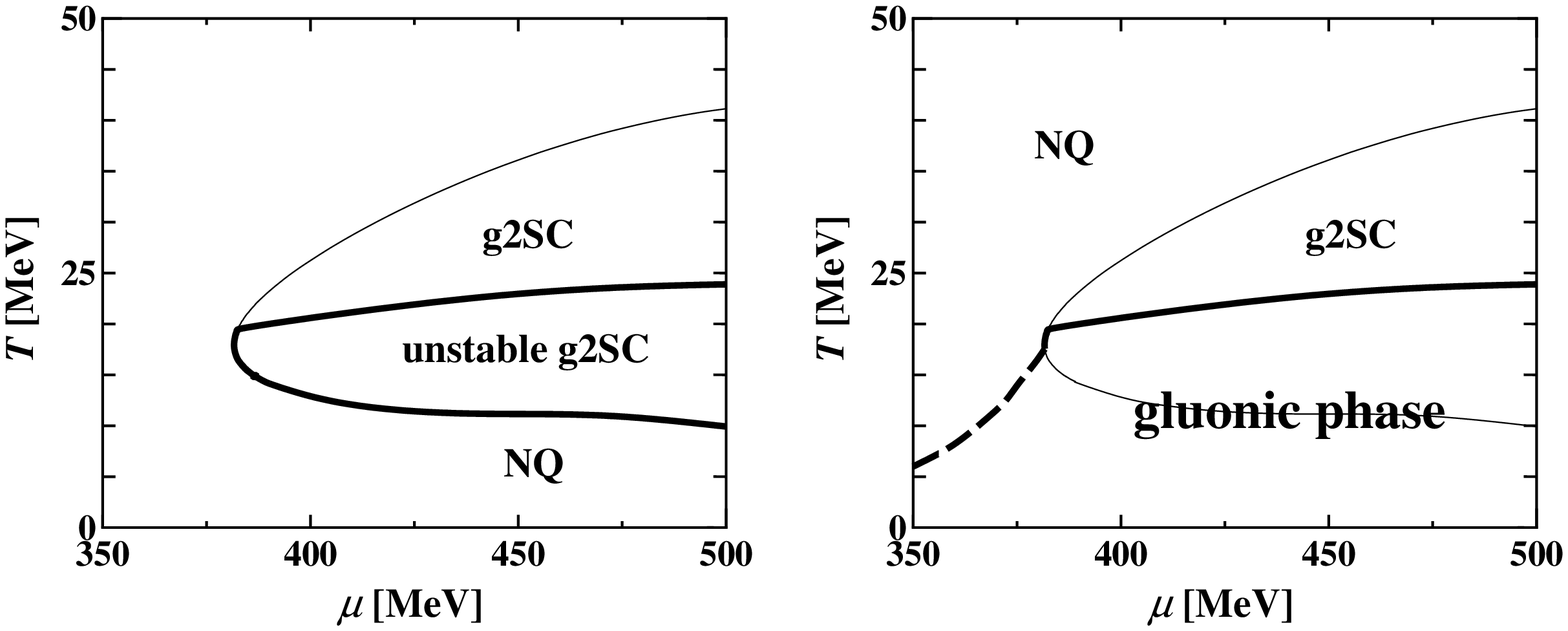}}
\caption{Left: The phase diagram of neutral three-flavor quark matter
in $T$-$\mu$ plane, with the diquark coupling chosen so that
the 2SC gap $\Delta_0=80~{\rm MeV}$ at $T=0$, $\mu=400~{\rm MeV}$ 
and $M_s\to\infty$.
Note that only the 2SC/g2SC phases are included in the analysis.
In the region enclosed by the thick solid line,
the Meissner masses of gluons 4--7 are tachyonic. 
Right: The same as the left panel, but the gluonic phase
is taken into account.
The transition between the gluonic phase and the g2SC phase is of second order
or weakly first order. On the other hand, the gluonic phase undergoes
a strong first-order transition to the NQ phase.}
\label{Figure5}
\end{figure}

From the result shown in Fig. \ref{Figure4}, 
we could imagine that a large region of 
the 2SC/g2SC/NQ phases in the intermediately coupled cold 
dense quark matter are replaced by the gluonic phase. 
To be more explicit and realistic, we used a three-flavor 
gauged NJL model 
and investigated the 2SC/g2SC/NQ phases in the $T$-$\mu$ phase diagram 
(left panel of Fig. \ref{Figure5}). 
We used a (density-independent) fixed value 
of the strange quark mass $(M_s=500~{\rm MeV})$. 
The diquark coupling was chosen so that 
the 2SC gap $\Delta_0=80~{\rm MeV}$ at $T=0$, $\mu=400~{\rm MeV}$ 
and $M_s\to\infty$. 
It should be noted that only the 2SC/g2SC/NQ phases 
were included in this analysis. 
We also neglected the dynamical breakdown of chiral symmetry. 
Nevertheless, at least qualitatively, 
this phase diagram is consistent with those presented in Ref. \citen{phased}. 
We again studied the gluonic phase in a self-consistent manner 
and obtained a phase diagram displayed 
in the right panel of Fig. \ref{Figure5}. 
One can see that the unstable g2SC phase is replaced by 
the gluonic phase and, furthermore, the gluonic phase is favored over 
the stable NQ phase in the low-temperature regime. 

\section{Summary}
We studied the chromomagnetic instability in the 2SC/g2SC phases 
at moderate density and at finite temperature. 
We calculated the Meissner masses squared of gluons 4--7 and 8 
and mapped out the unstable regions on the phase diagram. 
In order to resolve the chromomagnetic instability, 
we investigated the phases with gluonic vector condensations, 
i.e., the gluonic phase and the single plane-wave LOFF phase. 
Using the gauged NJL model and the mean-field approximation, 
we computed the free energies of the gluonic/LOFF phases 
in a self-consistent manner and explored the phase structure:\\
\noindent
$\bullet$ The gluonic phase is favored over the LOFF phase 
in a wide range of coupling strength.\\
\noindent
$\bullet$ The effective potential shows a peculiar behavior 
as a function of the vector condensations. 
It is particularly interesting that, in the weak-coupling regime, 
the gluonic/LOFF phases could be energetically more favored 
than the chromomagnetically stable NQ phase.\\
\noindent
$\bullet$ The strongly first-order transition between the gluonic phase 
and the NQ phase takes place. 
On the other hand, the transition from the gluonic phase 
to the stable g2SC is of second order or weakly first-order.

In addition to the two-flavor case, we also explored the 2SC/g2SC 
and the gluonic phases in neutral three-flavor quark matter. 
The resulting $T$-$\mu$ phase diagram shows clearly that 
currently known phase diagrams must be significantly altered. 

As we have shown, the gluonic phase is energetically more favored 
than the unstable 2SC/g2SC phases. 
However, the project of finding the true ground state of 
neutral dense quark matter is still not complete. 
In order to draw a conclusion, we have to investigate LOFF phases 
with multiple plane waves 
and other types of the gluonic phase.\cite{RS,Gorbar2005} 
It is also indispensable to look at the chromomagnetic stability 
of those phases.\cite{HJ} 

Although the result obtained from 
the (gauged) NJL models seems to be qualitatively reasonable, 
it should be unreliable quantitatively. 
Therefore, a novel technique is required 
to perform nonperturbative QCD calculations at relevant densities 
and to obtain a plausible QCD phase diagram.

\section*{Acknowledgements}
I thank Dirk Rischke and Igor Shovkovy for fruitful discussions. 
I also thank the organizers of YITP international symposium 
on ``Fundamental Problems in Hot and/or Dense QCD", 
for their invitation and warm hospitality. 
This work was supported by the Deutsche Forschungsgemeinschaft (DFG).

\end{document}